\documentclass[
    ,final            
    ,sort&compress
	]
  {aipproc}

\layoutstyle{8x11double}

\usepackage{amsmath}
\begin{document}

\title{Light-meson properties from the Bethe-Salpeter equation}


\author{M. Blank}{
  address={Institut f\"ur Physik, Karl-Franzens-Universit\"at Graz, A-8010 Graz, Austria}
}

\author{A. Krassnigg}{
  address={Institut f\"ur Physik, Karl-Franzens-Universit\"at Graz, A-8010 Graz, Austria}
}

\begin{abstract}
We discuss how to extract observables from an inhomogeneous vertex
Bethe-Salpeter equation without
resorting to the corresponding homogeneous equation. As an example we
present a prediction for
the $e^+e^-$ decay width of the $\rho(1450)$ or $\rho'$ meson. We also
attempt to identify the
momentum range contributing to a vector meson's decay constant.
\end{abstract}

\maketitle


\section{Introduction}

Due to their quantum numbers, vector-meson resonances are easily
produced in $e^+e^-$ scattering.
In addition, the $\rho$ is among the lightest mesons and thus object of
numerous studies and
a prime target for theoretical investigation. In QCD, the
Dyson-Schwinger-equation approach 
offers a nonperturbative continuum
method to study mesons
as bound states of quarks and gluons via the Bethe-Salpeter equation.
Herein we present some new results in the light of recent progress
regarding the methods used
to approach such a bound-state problem.

\section{The vertex BSE}
A general vertex $\Gamma(q,P)$, the inhomogeneous Bethe-Salpeter amplitude (iBSA), that connects quark and anti-quark to a color-singlet current satisfies the equation
\begin{equation}\label{eq:gammainh}
 \Gamma(p,P)=\Gamma_0+\int \frac{d^4 q}{(2 \pi)^4}K(P,p,q)S_a(q_+)\Gamma(q,P)S_b(q_-)
\end{equation}
(inhomogeneous or vertex BSE), where $K$ is the quark-antiquark scattering kernel, and $\Gamma_0$ a current which defines the channel under investigation by spin, parity and charge-conjugation parity. In this work, we investigate vector quantum numbers, $J^{PC}=1^{--}$, such that we choose (cf. \cite{Maris:1999bh})
\begin{equation}
\Gamma_0=Z_2 \gamma_\mu \;,
\end{equation}
where the renormalization constant $Z_2$ is calculated from the gap equation, cf. \cite{Maris:1997tm}.

The iBSA has poles at the positions of the respective bound states, and it can be decomposed as
\begin{equation}\label{eq:gammapole}
 \Gamma^\mu(q,P)=\sum_i\mathcal N_i\frac{\Gamma_{[h]}^\mu(q,P_i)}{P^2-P_i^2}\:+\:\mathrm{regular\; terms}\;,
\end{equation}
where $\Gamma_{[h]}$ is the homogeneous BSA of the meson under investigation, $P_i^2=-M_i^2$ the total momentum squared of the excitation $i$ in the respective channel, and $\mathcal N_i$ denotes a normalization constant.

We work in the well-established setup of the rainbow-ladder truncation and the effective interaction proposed by Maris and Tandy \cite{Maris:1999nt}, with light-quark masses and the parameter $\omega=0.4$ GeV (except where noted), as given in \cite{Krassnigg:2009zh}.

\section{Masses and decay constants}
According to Eq. (\ref{eq:gammapole}), each bound state results in a pole in the iBSA $\Gamma^\mu(P,q)$. As described in \cite{Blank:2010bp}, the amplitude is decomposed into components $F_i(P^2,q^2,P\cdot q)$ and covariants $T_\mu^i(P,q,\gamma)$ according to 
\begin{equation}
 \Gamma^\mu(P^2,q^2,P\cdot q)=\sum_i F_i(P^2,q^2,P\cdot q) T_\mu^i(P,q,\gamma)\;,
\end{equation}
where the standard vector covariants \cite{Krassnigg:2009zh} are orthonormalized according to $\mathrm{Tr}[T_\mu^i(P,q,\gamma)T_\mu^j(P,q,\gamma)]=\delta_{ij}$. It is well-known how to obtain meson masses and decay constants using the corresponding homogeneous BSE. Here, however, we explore a different approach.

\begin{figure}
\includegraphics[width=0.95\columnwidth]{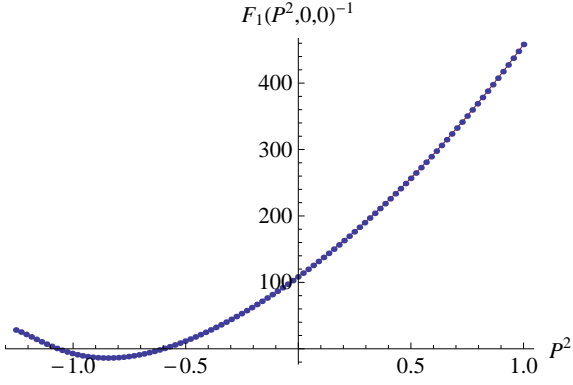}
\caption{\label{fig:massfit}The inverse of the first component of the inhomogeneous vector amplitude $1/F_1(P^2,0,0)$, as a function of the square of the total momentum $P^2$. The zero-crossings give the masses of the ground state and the first excitation.} 
\end{figure}

We calculate the bound state masses by fitting the inverse of the first component $F_1(P^2,0,0)$, as shown in Fig. \ref{fig:massfit}. We obtain
\begin{equation}\label{eq:mresults}
m_\rho=0.774\;\mathrm{GeV}\qquad
m_{\rho'}=1.034\;\mathrm{GeV}
\end{equation}
which is in agreement with \cite{Krassnigg:2009zh} and \cite{Krassnigg:2008gd}. 

The decay constant for a vector meson is given by \cite{Maris:1999nt}
\begin{equation}\label{eq:fvhom}
M_i f_v(P_i^2) = \frac{Z_2}{3} \int \frac{d^4 q}{(2 \pi)^4} \mathrm{Tr}[\gamma_\mu S_a(q_+)\Gamma_{[h]}^\mu(q,P_i)S_b(q_-)]\;,
\end{equation}
where the trace runs over color and Dirac-indices. To extract the same information from the iBSA, we first consider the general case of a projection $f^{(ih)}_{\tilde{\Gamma}}(P^2)$ of the iBSA $\Gamma(q,P)$ on a current $\tilde{\Gamma}$,
\begin{equation}\label{eq:inhomproj1}
f^{(ih)}_{\tilde{\Gamma}}(P^2)=\int \frac{d^4 q}{(2 \pi)^4} \mathrm{Tr}[\tilde{\Gamma}\: S_a(q_+)\Gamma(q,P)S_b(q_-)]\;.
\end{equation}
The poles in $\Gamma(q,P)$ translate into poles of $f^{(ih)}_{\tilde{\Gamma}}(P^2)$. In order to calculate the corresponding on-shell projection $f_{\tilde{\Gamma}}(P^2)$ (which ultimately gives the decay constant), the inhomogeneous BSE is rewritten as  \cite{Maris:1999bh}
\begin{equation}\label{eq:bserewrite}
\Gamma(p,P)=\Gamma_0 + \int \frac{d^4 q}{(2 \pi)^4} M(P,p,q) S_a(q_+)\Gamma_0 S_b(q_-).
\end{equation}
$M(P,p,q)$ denotes the fully amputated quark-antiquark scattering matrix which contains the bound state poles and may therefore be written as \cite{Maris:1999bh,Maris:1997hd}
\begin{equation}\label{eq:m}
 M(P,p,q)=\sum_i\frac{\Gamma_{[h]}(p,P_i)\bar{\Gamma}_{[h]}(q,-P_i)}{P^2-P_i^2}\:+\:\mathrm{reg.\; terms}\;,
\end{equation}
with the index $i$ labeling the bound states in this channel, and $\bar{\Gamma}_{[h]}$ denoting the charge conjugation of $\Gamma_{[h]}$ (see, e.g., \cite{Maris:1997tm}).
Inserting Eqs. (\ref{eq:bserewrite}) and (\ref{eq:m}) in Eq. (\ref{eq:inhomproj1}), we find
\begin{multline}\label{eq:inhomproj}
f^{(ih)}_{\tilde{\Gamma}}(P^2)=\\
 \sum_i\int \frac{d^4 p\;d^4q}{(2 \pi)^8}\:\mathrm{Tr}[\tilde\Gamma\: S_a(p_+)\Gamma_{[h]}(p,P_i)S_b(p_-)]\\
\times\frac{\mathrm{Tr}[\bar{\Gamma}_{[h]}(q,-P_i) S_a(q_+)\Gamma_0 S_b(q_-)]}{P^2-P_i^2}\:+\:\mathrm{reg.\; terms}\;.
 \end{multline}
Thus, in order to obtain $f_{\tilde{\Gamma}}(P_i^2)$, one has to divide the residue of $f^{(ih)}_{\tilde{\Gamma}}(P^2)$ by the square root of the corresponding residue of the projection on the inhomogeneous term, $f^{(ih)}_{\Gamma_0}(P^2)$. 

\begin{figure}
\includegraphics[width=0.95\columnwidth]{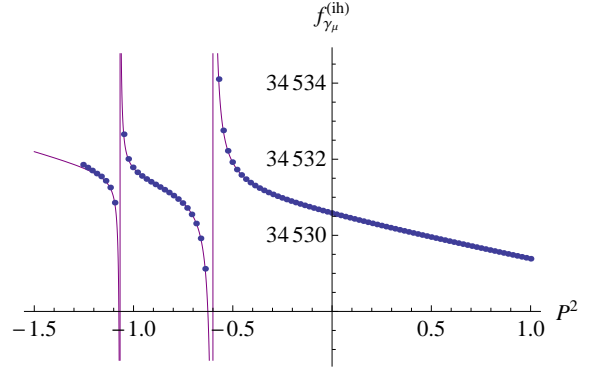}
\caption{\label{fig:fvinhom}The projection $f_{\gamma_\mu}^{(ih)}(P^2)$ defined in Eq. (\ref{eq:inhomproj1}). The line represents a pole fit to the data points, where the pole positions were taken from the fit of Fig. \ref{fig:massfit}.} 
\end{figure}

In the case of $f_v(P_i^2)$ the inhomogeneous term and the current are identical, $\Gamma_0\equiv\tilde\Gamma=Z_2\gamma_\mu$. The decay constants can therefore be calculated using only the residues $r_i$ of $f^{(ih)}_{\gamma_\mu}(P^2)$, which are extracted from the pole fit shown in Fig. \ref{fig:fvinhom}, where the masses (\ref{eq:mresults}) are used as input. Comparing Eqs. (\ref{eq:inhomproj}) and (\ref{eq:fvhom}), we find
\begin{equation}
 f_v(P_i^2)=\sqrt{\frac{r_i}{3 (-P_i^2)}}\;.
\end{equation}

\section{Prediction for $f_{\rho'}$}

The resulting decay constants of the $\rho$-meson and its first radial
excitation corresponding to the masses in (\ref{eq:mresults}) are
\begin{equation}\label{eq:fresults}
f_\rho=0.213\;\mathrm{GeV}\qquad
f_{\rho'}=0.079\;\mathrm{GeV}\;,
\end{equation}
which perfectly agrees with the results from the corresponding homogeneous BSE. To arrive at a concrete prediction for the decay width of the $\rho'$ into $e^+e^-$ we need to investigate the sensitivity of these results to the characteristic parameter of the model of Ref.~\cite{Maris:1999nt}. Indeed, the excited-state result exhibits a considerable dependence, while for the ground state the results for both $m$ and $f$ are rather solid, i.e., a dependence is observed but small (of the order of 2\% for $m$ and 7\% for $f$, see  
\cite{Holl:2004fr,Holl:2005vu,Krassnigg:2009zh} for in-depth information and a thorough discussion)

Here the important point is that one can use the systematic behavior of this dependence to determine both a preferred parameter
value or range for an ideal description in a phenomenological sense as well as to provide an estimate of the systematic effects in the calculation. As `best parameter' in our case we extract $\omega=0.5$ GeV and obtain
\begin{equation}\label{eq:groundresults}
m_\rho=0.762\;\mathrm{GeV}\qquad
f_{\rho}=0.218\;\mathrm{GeV}\;,
\end{equation}
for the ground state, which translates into a width $\Gamma_{e^+e^-}$ of
$6.92$ keV. The PDG \cite{Nakamura:2010zzi}
quotes an experimental value of $7.04\pm 0.06$ keV. For the $\rho'$ the
experimental situation is less clear: The PDG quote but don't use two
results of the order of and smaller than $0.1$ keV. Our excited-state
result corresponding to (\ref{eq:groundresults}) lies in a range accessible to us only via
extrapolation techniques due to the analytic
structure of the quark propagator (see, e.g., \cite{Bhagwat:2002tx} for
a discussion). However, we use extrapolations both
in model-parameter and momentum space to reduce the uncertainty.
We find for $\rho'$
\begin{equation}\label{eq:excitedresults}
f_{\rho'}= 0.095\pm 0.039\;\mathrm{GeV}   \quad  \Gamma_{e^+e^-}=0.94\pm0.66\;\mathrm{keV}\;.
\end{equation}

\section{Disassembling $f_\rho$}
\begin{figure}
\includegraphics[width=0.95\columnwidth]{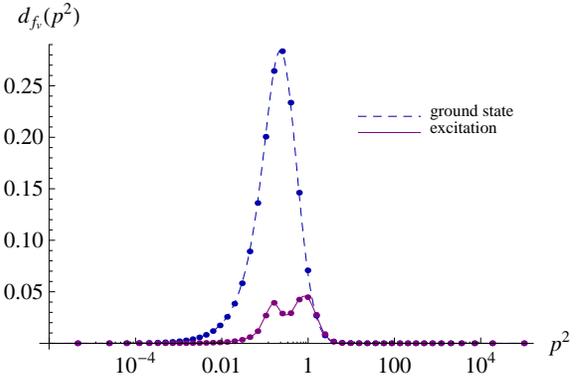}
\caption{\label{fig:fvpsq} Plot of the density $d_{f_v}(p^2)$ for the vector ground and excited states.} 
\end{figure}

To obtain further information on the structure of the $\rho$-meson and its radial excitation, the contributions to the decay constants from the different momentum scales are investigated. Therefore, we solve in addition to the inhomogeneous also the homogeneous BSE, from which the decay constant can be calculated according to Eq. (\ref{eq:fvhom}), and define the density $d_{f_v}(p^2)$ via
\begin{equation}
f_v(P_i^2) = \left(1/\sqrt{-P_i^2}\right) \int dp^2\:d_{f_v}(p^2)\;.
\end{equation}
$d_{f_v}(p^2)$ is plotted in Fig. \ref{fig:fvpsq} for the ground state and first radial excitation of the $\rho$-meson. The main contributions are centered in the mid-momentum regime around $p^2=0.1\;\mathrm{GeV}^2$, and neither the UV nor the IR have a strong influence. This supports \cite{Blank:2010pa}, where it was shown that $f_\rho$ (among other quantities) is insensitive to the behavior of the effective interaction in the far infrared. The same is true for the (perturbatively determined) UV domain. Therefore, an effective interaction, which neglects this, quite naturally gives reliable results for the decay constant as well, cf. \cite{Alkofer:2002bp}.

We further note that the scale dependence of analogous projections has also been investigated in lattice QCD \cite{Glozman:2010vd}.

\section{Conclusions}
We have demonstrated how the masses and decay constants of the $\rho$ meson and its radial excitation can be calculated from the inhomogeneous (vertex) Bethe-Salpeter equation without using any information from the corresponding homogeneous solutions, which is in contrast to a previous study \cite{Bhagwat:2007rj}. We have exemplified our method via a reasonable result for $f_{\rho'}$.

Furthermore, we have investigated the contributions to $f_\rho$ and $f_{\rho'}$ from different momentum scales, and found significant contributions from neither the far IR nor the UV domain.


\begin{theacknowledgments}
We would like to acknowledge valuable discussions with R.~Alkofer, G.~Eichmann, D.~Horvati\'c, C.~B.~Lang, M.~Limmer, and V.~Mader. This work was supported by the Austrian Science Fund \emph{FWF} under project no.\ P20496-N16,  and
was performed in association with and supported in part by the \emph{FWF} doctoral program no.\ W1203-N08.
\end{theacknowledgments}



\bibliographystyle{aipproc-mod}   


\begin{thebibliography}{15}
\expandafter\ifx\csname natexlab\endcsname\relax\def\natexlab#1{#1}\fi
\providecommand{\enquote}[1]{``#1''}
\expandafter\ifx\csname url\endcsname\relax
  \def\url#1{\texttt{#1}}\fi
\expandafter\ifx\csname urlprefix\endcsname\relax\def\urlprefix{URL }\fi
\providecommand{\eprint}[2][]{#1\url{#2}}

\bibitem[Maris and Tandy(2000)]{Maris:1999bh}
P.~Maris, and P.~C. Tandy, \emph{Phys. Rev. C} \textbf{61}, 045202 (2000).

\bibitem[Maris and Roberts(1997)]{Maris:1997tm}
P.~Maris, and C.~D. Roberts, \emph{Phys. Rev. C} \textbf{56}, 3369--3383
  (1997).

\bibitem[Maris and Tandy(1999)]{Maris:1999nt}
P.~Maris, and P.~C. Tandy, \emph{Phys. Rev. C} \textbf{60}, 055214 (1999).

\bibitem[Krassnigg(2009{\natexlab{a}})]{Krassnigg:2009zh}
A.~Krassnigg, \emph{Phys. Rev. D} \textbf{80}, 114010 (2009{\natexlab{a}}).

\bibitem[Blank and Krassnigg(2010)]{Blank:2010bp}
M.~Blank, and A.~Krassnigg  (2010), \eprint[arXiv:]{1009.1535}.

\bibitem[Krassnigg(2009{\natexlab{b}})]{Krassnigg:2008gd}
A.~Krassnigg, \emph{PoS} \textbf{Confinement8}, 75 (2009{\natexlab{b}}).

\bibitem[Maris et~al.(1998)]{Maris:1997hd}
P.~Maris, C.~D. Roberts, and P.~C. Tandy, \emph{Phys. Lett. B} \textbf{420},
  267--273 (1998).

\bibitem[Holl et~al.(2004)]{Holl:2004fr}
A.~H\"oll, A.~Krassnigg, and C.~D. Roberts, \emph{Phys. Rev. C} \textbf{70},
  042203(R) (2004).

\bibitem[Holl et~al.(2005)]{Holl:2005vu}
A.~H\"oll, A.~Krassnigg, P.~Maris, C.~D. Roberts, and S.~V. Wright, \emph{Phys.
  Rev. C} \textbf{71}, 065204 (2005).

\bibitem[Nakamura et~al.(2010)]{Nakamura:2010zzi}
K.~Nakamura, et~al., \emph{J. Phys. G} \textbf{37}, 075021 (2010).

\bibitem[Bhagwat et~al.(2003)]{Bhagwat:2002tx}
M.~Bhagwat, M.~A. Pichowsky, and P.~C. Tandy, \emph{Phys. Rev. D} \textbf{67},
  054019 (2003).

\bibitem[Blank et~al.(2010)]{Blank:2010pa}
M.~Blank, A.~Krassnigg, and A.~Maas  (2010), \eprint[arXiv:]{1007.3901}.

\bibitem[Alkofer et~al.(2002)]{Alkofer:2002bp}
R.~Alkofer, P.~Watson, and H.~Weigel, \emph{Phys. Rev. D} \textbf{65}, 094026
  (2002).

\bibitem[Glozman et~al.(2010)]{Glozman:2010vd}
L.~Y. Glozman, C.~B. Lang, and M.~Limmer, \emph{PoS} \textbf{LATTICE2010}, 149
  (2010).

\bibitem[Bhagwat et~al.(2007)]{Bhagwat:2007rj}
M.~S. Bhagwat, A.~H\"oll, A.~Krassnigg, C.~D. Roberts, and S.~V. Wright,
  \emph{Few-Body Syst.} \textbf{40}, 209--235 (2007).

\end{thebibliography}

\end{document}